*Баліка С. Д.*

*Одеський національний університет імені І. І. Мечникова,*
*65026, вул. Дворянська, 2,Одеса, Україна*
*E-mail: svitlana.balika@onu.edu.ua*

**Вплив застійного шару на довжину вільного пробігу фотонів у концентрованих суспензіях наночастинок**

*У роботі аналізується можливість визначення товщини і показника заломлення застійного шару наночастинок у концентрованих суспензіях за транспортними характеристиками фотонів розсіяного світла Аналіз базується на фізично прозорому узагальненні поняття однократного розсіяння на системах, у яких кількість частинок в об'ємах з лінійними розмірами порядку довжини світлової хвилі в середовищі значно перевищує одиницю. Це узагальнення здійснюється в рамках уявлення про компактні групи частинок, дозволяє вийти за межі традиційного наближення Борна та врахувати багаточастинкові ефекти, яким відповідають ті області інтегрування в членах ітераційного ряду для розсіяного поля, де внутрішні пропагатори мають поведінку типу дельта-функції. Як результат, обчислення транспортних характеристик фотонів виявляється можливим без детального моделювання процесів багаточастинкових розсіянь і кореляцій у системі.*

*Досліджено теоретичну залежність довжини вільного пробігу фотонів від показника заломлення та товщини застійного шару, показано їх помітний вплив на неї. Збільшення показника заломлення при фіксованій товщині шару зменшує довжину пробігу внаслідок збільшення оптичної густини суспензії. Характер залежності довжини пробігу фотонів від товщини шару визначається співвідношенням між значеннями його показника заломлення і показника заломлення базової рідини. Вона є зростаючою, коли перший є меншим за другий, але спадає в противному випадку. Експериментально спостережуване збільшення довжини пробігу з концентрацією частинок традиційно пояснюється проявом вищих кореляційних ефектів. Наша теорія показує що наявність застійного шару заплутує ситуацію, оскільки обидва фактори можуть як підсилювати, так і послаблювати один одного. Для розв'язання цього питання потрібна постановка нових спеціально спланованих експериментів.*

***Ключові слова:*** *суспензії наночастинок, застійний шар, розсіяння світла, фотонний транспорт, гідродинамічний радіус*

**Вступ**

У нашому недавньому дослідженні [1] розглядається проблема вимірювання ζ-потенціалу нанофлюїдів на базі розчинів електролітів за допомогою одночасних вимірювань методами електричної спектроскопії та лазерної кореляційної спектроскопії. У рамках теорії [1] точність вимірювання потенціалу суттєвим чином залежить від значення товщини $u^*$ області між поверхнею частинки та площиною гідродинамічного зсуву, тобто товщини застійного шару. Теорія [1] ґрунтується на методі компактних груп неоднорідностей [2], застосованому до систем частинок з морфологією тверде ядро-проникна оболонка [3]. Знайдено аналітичну залежність провідності суспензії від ζ-потенціалу і $u^*$ для різних граничних значень провідності застійного шару. Показано, що в цих випадках знання тангенса кута нахилу концентраційної залежності провідності нанофлюїдів $\partial \sigma_{eff} / \partial c$, віднесеного до провідності базової рідини $\sigma_0$, дозволяє встановити однозначну залежність ζ-потенціалу від $u^*$. Тим самим задача визначення ζ-потенціалу фактично зводиться до визначення величини $\sigma_0^{-1} \partial \sigma_{eff} / \partial c$ та гідродинамічного радіус частинки, наприклад, через коефіцієнт дифузії Смолуховського-Ейнштейна.

Слід зауважити, що поняття фізичного розміру частинки в електроліті відрізняється від звичайного геометричного розміру частинки, оскільки навколо частинки формується подвійний електричний шар (ПЕШ). Проблема додатково ускладнюється тим фактом, що сам ПЕШ має складну мікроструктуру, що включає шар Штерна, застійну частину дифузного ПЕШ та мобільну частину дифузного ПЕШ. В залежності від розчину та характеристик системи мобільна дифузна частина ПЕШ може бути сильно вираженою (суспензії на базі помірно концентрованих водних розчинів електролітів) або навпаки, подавленою (суспензії на базі концентрованих водних розчинів електролітів або на базі неводних розчинів). Враховуючи, що характерна товщина шару Штерна має розмір іона, природно припустити, що фізичний розмір наночастинок у суспензіях другого типу фактично формується завдяки існуванню застійного шару.

Методика вимірювання фізичного радіуса частинок методом кореляційної спектроскопії було продемонстрована в роботі [4] для нанофлюїдів ізопропанол/$Al_2O_3$. Згідно з цими даними фізичний радіус досить швидко зростає (до 60%) навіть при помірних варіаціях концентрації наночастинок. Фактично це означає, що коефіцієнт дифузії частинки, через який фізичний радіус частинки визначався за допомогою формули Смолухоського-Ейнштейна, помітно зменшується при таких варіаціях концентрації. На наш погляд, цей факт не зовсім зрозумілий, оскільки можна було б очікувати, що властивості застійного шару, зокрема, його товщина, у першу чергу визначаються характером міжмолекулярних взаємодій на поверхні розділу частинка-рідина, а вплив інших наночастинок на фізичний розмір виділеної частинки повинен був би проявлятися лише при достатньо щільному їх упакуванні. Як альтернативне пояснення, у цій роботі ми припускаємо, що спостережуване зменшення коефіцієнта дифузії може бути пов'язане з тим, що припущення про незалежність броунівського руху окремих частинок починає порушуватися при зростанні концентрації наночастинок (рух частинок перешкоджається іншими частинками), починає проявлятися гідродинамічна взаємодія частинок тощо. У зв'язку з цим виникає проблема пошуку інших незалежних методів оцінки гідродинамічного радіуса частинки. У даній роботі аналізується можливість визначення товщини застійного шару за транспортними характеристиками поширення фотонів [5] у концентрованих суспензіях наночастинок. Аналіз базується на роботах [6,7], де в рамках ідеї про розсіяння на компактних групах розсіювальних центрів вводиться поняття про однократне розсіяння в узагальненому розумінні на концентрованих системах, де кількість частинок в об'ємі $\lambda^3$ ($\lambda$ – довжина світлової хвилі в середовищі) значно перевищує одиницю. Відповідний аналіз транспортних характеристик виявляється можливим [7] без використання надмірної деталізації процесів розсіяння в системі. Деякі технічні деталі і попередні результати нашого аналізу були представлені раніше [8].

**1. Модель системи і знаходження інтенсивності розсіяння світла**

Для простоти розглядаємо суспензії із сильно подавленою мобільною частиною ПЕШ. Кожна частинка суспензії і прилеглий до неї застійний шар моделюємо як тверде ядро з радіусом $R_1$, оточене твердим концентричним сферичним шаром із зовнішнім радіусом $R_2$ (рис. 1). Частинки разом із застійними шарами дисперговані в однорідне середовище з показником заломлення $n_0$. Показник заломлення ядра – $n_1$, оболонки – $n_2$. Показники заломлення вказаних структурних одиниць пов'язані з діелектричними проникностями цих одиниць співвідношеннями виду $n = \sqrt{\varepsilon}$.

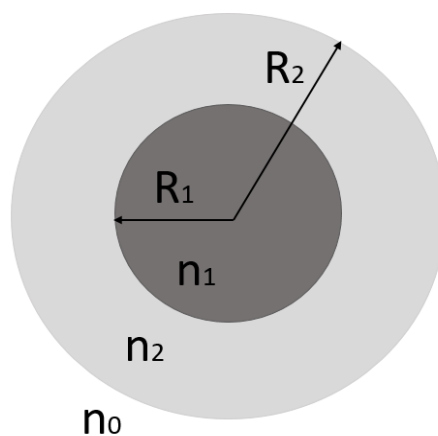

**Рис.1.** Модель твердої частинки радіусом $R_1$, оточеної застійним шаром радіусом $R_2$, диспергованої в однорідне середовище з показником заломлення $n_0$. Показник заломлення ядра - $n_1$, оболонки – $n_2$.

Ми вважаємо, що об'ємна концентрація частинок $c$ у суспензіях достатньо висока. Елементарні оцінки показують, що кількість нанорозмірних частинок радіусом $R$ в одиниці об'єму суспензії становить $n = 3c/(4\pi R^3)$, і відповідно кількість наночастинок в об'ємі $\lambda^3$ (групи частинок усередині таких об'ємів називатимемо компактними) становить $N = n\lambda^3 = \dfrac{3c}{4\pi}\left(\dfrac{\lambda}{R}\right)^3$. Зокрема, для частинок з $R = 50$ нм і видимого світла з $\lambda = 500$ нм $N \approx 240\,c$, тобто вже при $c \geq 0.04$ кількість таких наночастинок в об'ємі $\lambda^3$ на порядок перевищує одиницю. Цей факт дозволяє нам для обчислення транспортних характеристик фотонів в таких суспензіях безпосередньо скористатися підходом [7].

З фізичної точки зору, розсіяння на вказаних компактних групах наночастинок частинок є однократним. Згідно з теорією [6] локальний розподіл діелектричної проникності в нашій модельній системі записується у вигляді

$$\varepsilon(\mathbf{r}) = \varepsilon_0 + \delta\varepsilon(\mathbf{r}) \qquad (1)$$

де $\delta\varepsilon(\mathbf{r})$ – внесок компактної групи. У нашому випадку

$$\delta\varepsilon(\mathbf{r}) = \sum_{i=1}^{N}\left\{\Delta\varepsilon_1 \theta(R_1 - |\mathbf{r} - \mathbf{r}_i|) + \Delta\varepsilon_2[\theta(R_2 - |\mathbf{r} - \mathbf{r}_i|) - \theta(R_1 - |\mathbf{r} - \mathbf{r}_i|)]\right\} \qquad (2)$$

де $\Delta\varepsilon_1(\mathbf{r}) = \varepsilon_1(\mathbf{r}) - \varepsilon_0, \Delta\varepsilon_2(\mathbf{r}) = \varepsilon_2(\mathbf{r}) - \varepsilon_0$ і $\theta(x) = \begin{cases} 1, x > 0 \\ 0, x < 0 \end{cases}$ – ступінчата функція Хевісайда.

Класичний підхід до вивчення розсіяння світла рідинами далеко від критичної точки обмежується використанням першого наближення Борна в теорії багатократного розсіяння [5]. Це означає, що враховуються ефекти розсіяння, пов'язані лише з парами частинок. Це, очевидно, не зовсім правильно: наприклад, слід очікувати, що в рідинах взаємодії між сусідніми частинками поляризують одна одну більш інтенсивно, ніж у газах, але опис розсіяння в обох випадках здійснюється аналогічно. Метод компактних груп усуває цей недолік теорії. А саме, вводиться поняття однократного розсіяння в узагальненому розумінні, під яким розуміється розсіяння на парах компактних груп частинок і яке з фізичного погляду є однократним. Крім традиційного розсіяння на парах окремих частинок, таке розсіяння включає багаточастинкові ефекти розсіяння і поляризації всередині компактних груп. Як було раніше показано, для наночастинок і видимого світла ці ефекти повинні ставати помітними вже для об'ємних концентрацій $c \geq 0.04$, коли кількість наночастинок в об'ємі $\lambda^3$ може значно перевищувати одиницю.

Виокремлення ефектів багатократного розсіяння та кореляцій всередині компактних груп базується [6] на тому факті, що компактним групам відповідають ті області змінних інтегрування в ітераційному ряді для розсіяного поля, де внутрішні пропагатори електромагнітного поля $T_{\alpha\beta}$ виявляють сингулярну поведінку типу дельта функції. Використовуючи спеціальне представлення для $T_{\alpha\beta}$, вказані ефекти можна виділити з усіх членів ітераційного ряду для розсіяного поля та підсумувати для довільних різниць між діелектричними проникностями дисперсних частинок і дисперсійного середовища та без знання деталей ближніх багаточастинкових кореляцій.

У результаті загальний вираз для інтенсивності однократного розсіяння (в узагальненому розмінні) світла в концентрованій суспензії має вигляд [7]

$$I(\mathbf{q}) = \sum_{r,s=1}^{\infty} I_{rs}(\mathbf{q}) \qquad (3)$$

де

$$I_{rs}(\mathbf{q}) \propto \left(-\frac{1}{3\varepsilon_0}\right)^{r+s-2} \int_V d\mathbf{r} \langle (\delta\varepsilon(\mathbf{r}))^r (\delta\varepsilon(0))^s \rangle e^{-i\mathbf{q}\cdot\mathbf{r}} \qquad (3)$$

є внеском від $r$ і $s$ актів розсіяння всередині пари компактних груп, $\mathbf{q}$ – зміна хвильового вектора світла внаслідок розсіяння.

Для макроскопічно однорідної і ізотропної системи твердих куль з кусково-неперервним радіальним розподілом діелектричної проникності для інтенсивності розсіяння (на ненульовий кут $\theta$) дістаємо [7]:

$$I(\mathbf{q}) = I_0 V \frac{k_0^4 \sin^2 \gamma}{2\pi R_0^2} \varepsilon_0^2 |\alpha(\mathbf{q})|^2 n S(\mathbf{q}), \qquad (4)$$

де $I_0$ – інтенсивність падаючої хвилі, $V$ – розсіювальний об'єм, $k_0$ – величина хвильового вектора світла у вакуумі, $\gamma$ – кут між вектором поляризації падаючої хвилі та напрямком на точку спостереження на відстані $R_0$, $n$ – кількість частинок в одиниці об'єму, $S(\mathbf{q})$ – статичний структурний фактор системи куль, $\alpha(\mathbf{q})$ – ефективна поляризовність частинки разом із застійним шаром у системі, яка в нашому випадку має вигляд:

$$\alpha(q) = \frac{3}{4\pi} \int_{0 \leq r \leq R_1} d\mathbf{r} e^{-i\mathbf{q}\mathbf{r}} \frac{\varepsilon_1 - \varepsilon_0}{\varepsilon_1 + 2\varepsilon_0} + \frac{3}{4\pi} \int_{R_1 \leq r \leq R_2} d\mathbf{r} e^{-i\mathbf{q}\mathbf{r}} \frac{\varepsilon_2 - \varepsilon_0}{\varepsilon_2 + 2\varepsilon_0} =$$

$$= \frac{\varepsilon_1 - \varepsilon_0}{\varepsilon_1 + 2\varepsilon_0} \frac{3R_1}{q^2} \left(\frac{\sin q R_1}{q R_1} - \cos q R_1\right) \qquad (5)$$

$$+ \frac{\varepsilon_2 - \varepsilon_0}{\varepsilon_2 + 2\varepsilon_0} \left[\frac{3R_2}{q^2} \left(\frac{\sin q R_2}{q R_2} - \cos q R_2\right) - \frac{3R_1}{q^2} \left(\frac{\sin q R_1}{q R_1} - \cos q R_1\right)\right]$$

## 2. Довжина вільного пробігу фотонів

Однією з найважливіших характеристик фотонного транспорту в середовищі є довжина вільного пробігу фотонів $l$. Вона визначає ступінь послаблення інтенсивності світла при поширенні в середовищі і входить у закон Бугера $I = I_0 \exp(-\alpha x)$, де $I$ – інтенсивність світла в середовищі після проходженні в ньому відстані $x$, $I_0$ – інтенсивність світла при входженні в середовище, $\alpha$ – коефіцієнт поглинання. За умови, коли послаблення інтенсивності світла відбувається лише внаслідок процесів розсіяння на оптичних неоднорідностях (наприклад, частинках суспензії), $\alpha = 1/l$ [5].

Довжина вільного пробігу фотонів обчислюється за формулою:

$$\frac{1}{l} = \frac{k_0^4 \varepsilon_0^2}{2} \int_\Omega d\Omega \left(1 + \cos^2 \theta\right) |\alpha(q)|^2 n S(q), \qquad (6)$$

Для оцінок цієї величин скористаємося строгими результатами для структурного фактора системи твердих куль у наближенні Перкуса-Йєвіка [9]:

$$S(q) = \frac{1}{1 - nC(q)} \qquad (7)$$

де $d = 2R_2$ – зовнішній діаметр кулі з оболонкою, $c(q)$ – Фур'є-образ прямої кореляційної функції:

$$c(q) = \frac{4\pi d}{q^2} \left[ (\alpha + \beta + \delta)\left(\cos qd - \frac{\sin qd}{qd}\right) - \beta\left(\frac{\sin qd}{qd} + \frac{2\cos qd - 1}{(qd)^2}\right) \right.$$
$$\left. - \delta\left(3\frac{\sin qd}{qd} + 12\frac{\cos qd}{(qd)^2} - 24\frac{\sin qd}{(qd)^3} - 24\frac{\cos qd - 1}{(qd)^4}\right) \right] \qquad (8)$$

$$\alpha = \frac{(2\eta + 1)^2}{(1-\eta)^4}, \quad \beta = -\frac{6\eta(\eta/2 + 1)^2}{(1-\eta)^4}, \quad \delta = \frac{\alpha \eta}{2}, \quad \eta = \frac{4\pi R_2^3 n}{3} \qquad (9)$$

### 3. Теоретична оцінка залежності $l$ від параметрів застійного шару

Скористаємося формулою (6) для аналізу залежності довжини вільного пробігу фотонів від товщини застійного шару для латексних частинок радіусом $R = 50$ нм у воді. Показники заломлення латексних частинок $n_1 = 1.59$ і води $n_0 = 1.33$. Нехай довжина хвилі падаючого світла у вакуумі $\lambda_0 = 665$ нм, що відповідає довжині $\lambda = 500$ нм у воді. Розглянемо чотири можливі ситуації, коли показник заломлення застійного шару $n_2$ задовольняє співвідношення $n_0 < n_1 \approx n_2$, $n_0 < n_1 < n_2$, $n_0 < n_2 < n_1$ і $n_2 < n_0 < n_1$. На Рис.2. представлено графіки залежності відносної довжини вільного пробігу фотонів з урахуванням застійного шару $l/l_0$ ($l_0$ – довжина вільного пробігу фотонів без урахування застійного шару) від відносної товщини застійного шару $u^*$ для цих ситуацій. Рисунок 3 демонструє залежність $l/l_0$ від показника заломлення застійного шару $n_2$ при фіксованій товщині $u^*$.

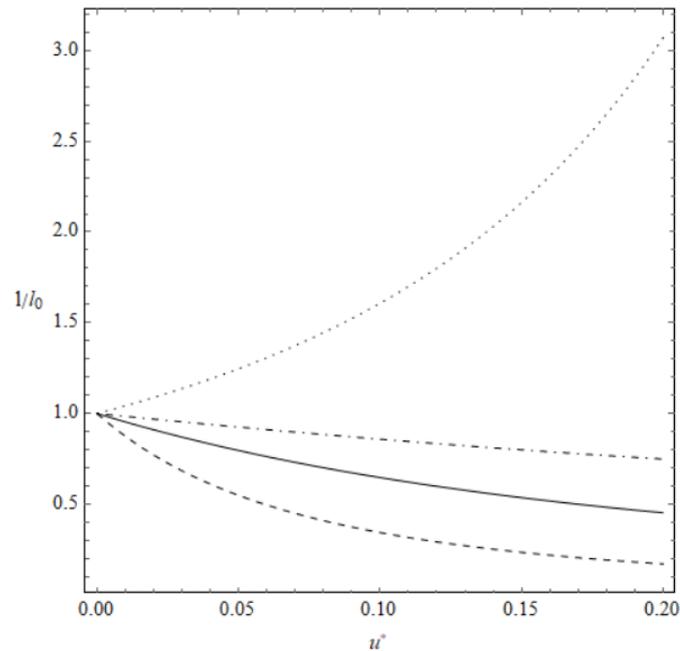

**Рис.2.** Теоретична залежність $l/l_0$ від $u^*$ для суспензій латексних частинок ($n_1 = 1.59$) радіусом $R = 50$ нм у воді ($n_0 = 1.33$). Суцільна крива – показник заломлення застійного шару дорівнює 1.6, штрихована – 2.09, штрих-пунктирна – 1.45, пунктирна – 1.2. Об'ємна концентрація частинок $c = 0.068$.

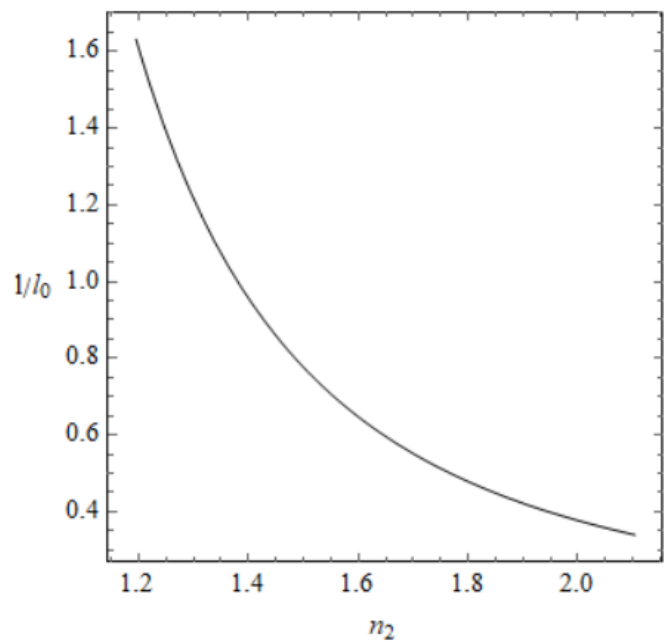

**Рис.3.** Теоретична залежність $l/l_0$ від показника заломлення застійного шару $n_2$ для суспензій латексних частинок ($n_1 = 1.59$) радіусом $R = 50$ нм і застійним шаром з відносною товщиною $u^* = 0.1$ у воді ($n_0 = 1.33$). Об'ємна концентрація частинок $c = 0.068$.

Результати обчислень показують, що $l$ виявляється чутливою як до значення показника заломлення шару $n_2$, так і до його товщини $u^*$. Однак характер цих залежностей дещо різниться. Зі збільшенням $n_2$ при фіксованих $n_0$ і $n_1$ довжина вільного пробігу фотонів зменшується. Фізично це зрозуміло, оскільки при такому збільшенні $n_2$ середовище стає оптично густішим, а тому ймовірність розсіяння фотонів зростає. Якісна ж залежність $l$ від товщини застійного шару при фіксованому значенні показника заломлення частинки може бути двох суттєво різних типів: спадною при зростанні $u^*$, якщо $n_2 > n_0$, і зростаючою при зростанні $u^*$, якщо $n_2 < n_0$. Ці типи залежностей відображають той факт, що формування застійного шару веде в першому випадку до збільшення оптичної густини середовища, яка посилюється зі зростанням $u^*$, тоді як у другому випадку оптична густина середовища зменшується.

Тенденція до збільшення $l$ порівняно зі значеннями, передбачуваними теорією Мі, спостерігалася в роботі [10], де вона пояснювалася внесками кореляційних ефектів вищих порядків між розсіювальними центрами в концентрованих суспензіях. Запропонована нами теорія не лише враховує ці ефекти, але й додатково вказує на ще один можливий механізм формування $l$, а саме, - формування застійного шару. Згідно з виконаними аналізом, внески від обох механізмів можуть як підсилювати один одного, так і конкурувати між собою. Для кількісного виокремлення цих внесків за допомогою запропонованої теорії потрібно мати достатньо екстенсивний набір експериментальних даних для суспензій з різними оптичними і геометричними характеристиками. На сьогодні подібні дані відсутні. Сподіваємося, що представлена робота стимулюватиме постановку відповідних експериментів.

# Balika S.D.

# The effect of the stagnant layer on the photon mean-free-path length in concentrated suspensions of nanoparticles

*We analyze the possibility of evaluation of the thickness and refractive index of the stagnant layer in concentrated suspensions of nanoparticles through the transport characteristics of scattered light photons. The analysis is based on a physically-transparent generalization of the concept of the single scattering intensity off systems in which the number of particles within regions with linear sizes of order of the wavelength in the medium greatly exceeds unity. This generalization is carried out within the notion of compact groups of particles, makes it possible to go beyond the traditional Born approximation, and take into account many-particle effects contributed from those ranges of integration variables in the terms of the iteration series for the scattered field where the internal propagators have delta-function-type behavior. As a result, the evaluation of the photon transport characteristics becomes possible without a detailed modeling of many-particle scattering and correlation processes in the system.*

*The photon mean-free-path length is investigated theoretically as a function of the stagnant refractive index and that of the layer thickness to show a noticeable effect of both parameters on it. As the layer refractive index is increased at a fixed layer thickness, the mean-free-path length decreases because the suspension optical density increases. As a function of the layer thickness, the photon mean-free-path length reveals different types of behavior, depending on the relation between the refractive indices of the stagnant layer and base liquid. If the former is smaller than the latter, this behavior is increasing; in the opposite case, it is decreasing. An experimentally observed increase of the photon mean-free-path length with the particle concentration is usually explained as a manifestation of higher correlation effects. Our theory reveals that the presence of the stagnant layer make the situation more complicated, for both factors may either enhance or diminish each other. To resolve the issue, new specially-designed experiments are required to be set up.*

***Keywords:*** *nanofluids, stagnant layer, light scattering, photon transport, hydrodynamic radius*